IAC-22-A7.3.4

# Design and Testing of a 3U CubeSat to Test the In-situ Vetoing for the vSOL Solar Neutrino Detector
## Jonathan Folkerts[a]

[a] *Department of Physics, Wichita State University, 1845 Fairmount St, Wichita, Kansas 67260, United States of America*, jdfolkerts@shockers.wichita.edu

### Abstract

For years, earth-based neutrino detectors have been run and operated to detect the elusive neutrino. These have historically been enormous underground detectors. The neutrino Solar Orbiting Laboratory (vSOL) project is working to design a technical demonstration to show that a much smaller neutrino detector can be operated in near-solar environments for a future spaceflight mission. At a closest approach of 3 solar radii, there is a ten thousand-fold increase in the neutrino flux. This would allow a 100 kg payload to be the equivalent of a 1 kTon earth-based payload, larger than the first neutrino experiment in the Homestake mine. As a continuing step towards this goal, the vSOL project will fly a 3U CubeSat for testing the detector's passive shielding design, active vetoing system in a space environment, and the rate of false double-pulse signals in a space environment. I go into technical detail about the characterization of the central detector in simuo and in the lab. The first test is a characterization of energy resolution and calibration through the use of radioactive sources. We will continue testing by measuring the veto success rate with ground-level cosmic rays. For the final ground testing, we will use the Fermilab test beam to characterize the central detector and veto performance at specific particle energies. Veto performance on the previous detector design has been promising, and we were able to veto a high percentage of all particles that can penetrate the passive shielding of the satellite. These laboratory results and simulations of the CubeSat detector design will raise the technological readiness level of the planned technological demonstrator flight to the sun, and the current level of shielding performance is promising for a successful CubeSat test flight.

## Acronyms/Abbreviations

Solar Radius (R$_\odot$)
Photomultiplier Tube (PMT)
Silicon Photomultiplier (SiPM)
Gallium Aluminum Gadolinium Garnet (GAGG)
Cerium-doped Gallium Aluminum Gadolinium Garnet (Ce:GAGG)
Geometry and Tracking 4 (Geant4)
Liquid Scintillator (LS)
Ultra-High frequency (UHF)
Printed Circuit Board (PCB)
Front-end Electronics (FEE)
Minimally Ionizing Particle (MIP)

## 1. Introduction

The vSOL project has been working since the first round of funding in 2016 to design and build a neutrino detector that is light and small enough to be placed into a near-solar orbit. Unlike other modern neutrino detectors, which gain their counting rates by having masses in the kiloton range and higher, this orbiting laboratory would gain increased counting rate by getting very close to the sun, as can be seen in Table 1.

Table 1. Solar neutrino flux relative to the Earth's flux at various points in the Solar System [1].

| Distance from Sun | Flux relative to Earth |
|---|---|
| 696342 km (R$_\odot$) | 46200. |
| 1500000 km (~3 R$_\odot$) | 10000. |
| 4700000 km (~7 R$_\odot$) | 1000. |
| 1500000 km | 100. |
| 47434000 km | 10. |
| Mercury | 6.4 |
| Venus | 1.9 |
| Earth (215 R$_\odot$) | 1 |
| Mars | 0.4 |
| Asteroid Belt | 0.1 |
| Jupiter | 0.037 |
| Saturn | 0.011 |
| Uranus | 0.0027 |
| Neptune | 0.00111 |
| Pluto | 0.00064 |
| KSP | 0.0002 |
| Voyager 1 (2015) | 0.00006 |

The other major challenge that vSOL faces relative to other neutrino detectors is the harsh radiation environment. Most neutrino detectors are built deep underground in mines, but vSOL will be exposed directly to the full spectrum of primary galactic cosmic ray backgrounds. The solution to this problem leads directly into the main thrust of this paper. The project proposes to use the double pulsed signal from the transmutation of gallium by a neutrino, summarized in (1), (2), and Figure 1.

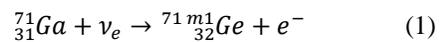

$$^{71}_{31}Ga + \nu_e \rightarrow {}^{71\,m1}_{32}Ge + e^- \qquad (1)$$

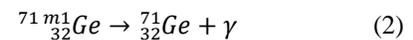

$$^{71\,m1}_{32}Ge \rightarrow {}^{71}_{32}Ge + \gamma \qquad (2)$$

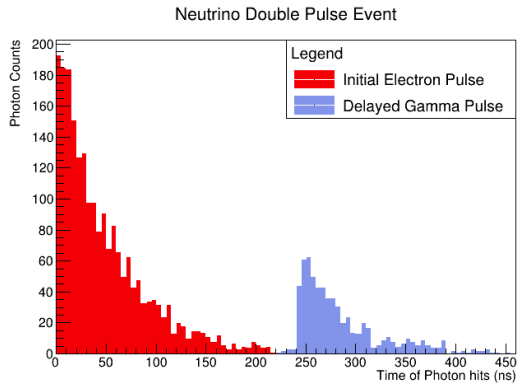

Fig. 1. Simulated double-pulse signal resulting from a neutrino interaction on gallium in a scintillator.

This interaction has a variety of options for characteristic timing to look for a double-pulse signal from the prompt electron and delayed gamma. This leads to the scientific motivation for the 3U CubeSat.

On the earth, it is impossible to directly measure the fake double-pulsing due to galactic cosmic rays due to the shielding from the atmosphere and the Van Allen belts. By operating a CubeSat in a polar orbit, we are able to spend approximately seven minutes every hour outside the radiation belts while above the north or south pole.

While outside the radiation belts, we will be able to take data of the raw cosmic ray spectrum without interference and make a measurement of the double-pulse false positive rate due to such a background. The secondary purpose of the CubeSat is the first test flight of a high-density re-entry safe shield constructed out of tungsten-loaded epoxy.

Of note to this paper is a design test from the NIAC phase II test of a veto array system. The first concept of the veto array used a series of PMTs to test our ability to consistently veto cosmic ray shower signals with a liquid scintillator inner detection volume. Due to their increasing capacity, lowering cost per cm², and some design constraints, we are considering completely replacing PMTs with SiPMs in our CubeSat test flight. The design and methodology to test such a change will be elaborated on in section 2.1.

## 2. Material and methods
### 2.1 CubeSat Design

The CubeSat bus and controlling equipment is being purchased commercially from NanoAvionics. The specifications of each component are proprietary and protected by a non-disclosure agreement, but the pieces are otherwise off-the-shelf. The components being used in the non-science payload are an S band antenna for regular communication, a UHF antenna for omnidirectional communication during detumbling and for emergencies, batteries and solar panels for power, and a control board. Most CubeSats of similar size use reaction wheels for pointing, but we only use magnetorquers. Other satellites typically fly cameras which require high pointing accuracy, but our science payload is designed to be operated isotopically, and the S band antenna only requires 90 degree pointing.

The science payload will consist of three pieces. The innermost piece, Figures 2 and 3, is an array consisting of cerium-doped Gallium Aluminum Gadolinium Garnet scintillating crystals. Each crystal is a 10 cm cube, and the four cubes are bonded together with an optical resin. This array is bonded to a 20 cm x 20 cm x 10 cm quartz optical window. The array of crystals is surrounded with a reflective coating and a thin plastic jacket for structure. The quartz viewing window is bonded to a SiPM array, Figure 4, which creates the inner detection volume. The signal from this device will be sent to a board for signal conditioning.

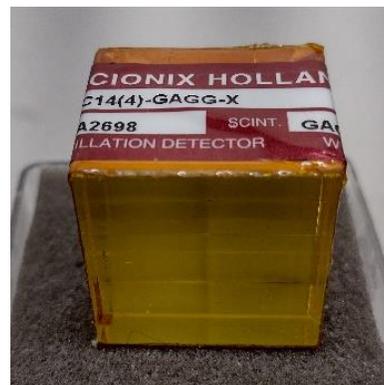

Fig. 2. GAGG scintillating crystal with quartz viewing window.

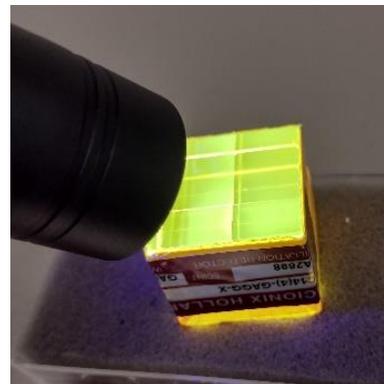

Fig. 3. Scintillating GAGG under UV light with characteristic yellow-orange glow

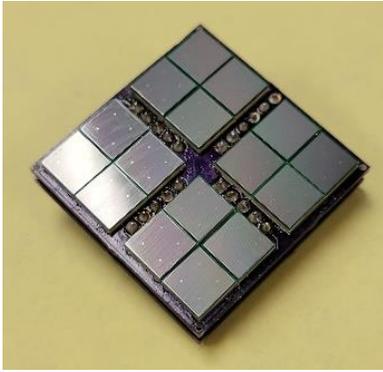

Fig. 4. SiPM array on PCB for GAGG crystal

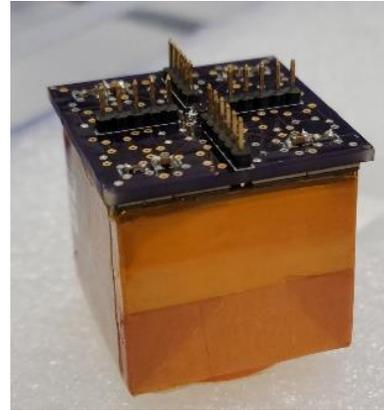

Fig. 5. SiPM array mounted on GAGG assembly

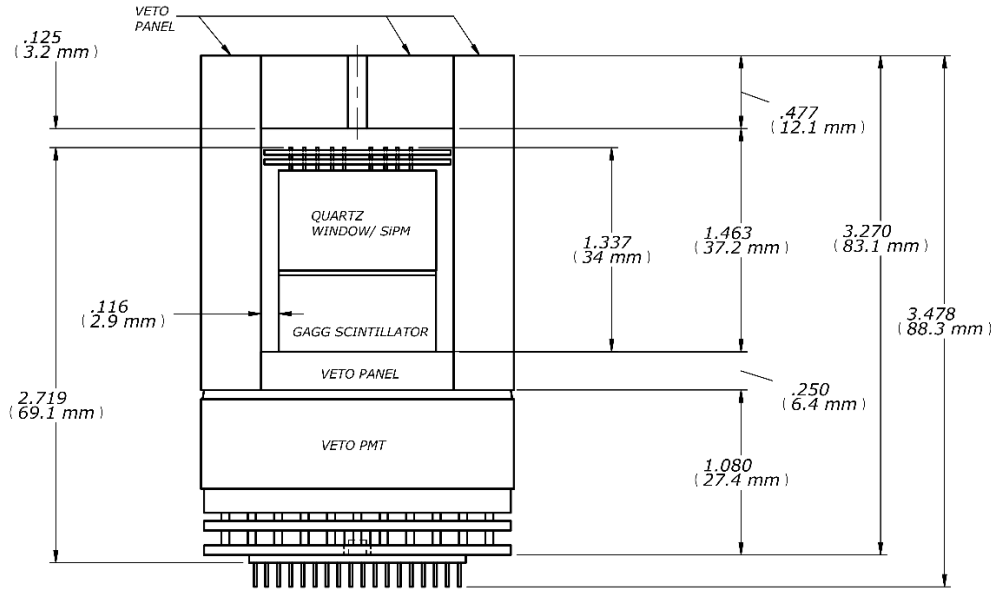

Fig. 6. Schematic of Veto Array and inner GAGG detector volume

The next layer of the science payload will be a veto array constructed of a series of plates of an Eljin plastic scintillator. These scintillators are to be covered with a reflective coating, and one end will be connected to either a PMT or to another SiPM array. This signal is sent to a FEE conditioning board for creating of a Boolean veto signal. Before final closing, the GAGG inner payload will be placed inside the veto with small openings made for the power and signal cables to run out.

The outermost layer of the science payload consists of a 1.5 cm thick high-density tungsten-loaded epoxy shield. The shield was created by mixing very small (0.3 − 20 μm) tungsten power in 3M Scotch-Weld 2216 epoxy resin. This mixture was poured into a silicone mold of the appropriate dimensions along with several aluminum pieces for mounting the shield to the CubeSat's frame. Before final closing, the veto array will be placed inside the shield, and then the shield will be closed using a small mixture of doped epoxy with as close to matching density as possible. Like the veto array, there will be small openings created during the final closing in order to run power in and signals out.

*2.2 Detector Characterization*

The characterization of the detector's performance has come in two major thrusts. First, we have operated and simulated the GAGG volume described in section 2.1 using a variety of radioactive sources including sodium 22, cesium 137, and barium 133. The detector's simulations were performed using Geant4, a c++ extension library for particle physics simulations [2], and the detector's operation was done using a Tektronix 3 series digital oscilloscope, DC power supplies, and small test radioactive sources.

Second, we have tested the active vetoing using an older detector geometry using a liquid scintillator inner volume. To test this system, we have used both PMTs and SiPMs as the detectors on the veto array to look for any differences between the two measurement methods. For this system, we have used the same digital oscilloscope, but, instead of radioactive sources, we use cosmic ray showers as our signal.

### 2.2.1 Initial Veto Array Design and Characterization

As part of the NIAC phase II grant, the vSOL collaboration tested an active veto array using PMTs mounted to a large plastic scintillating volume with an inner volume of liquid scintillator from the NOvA experiment, also read out by a PMT. The LS volume and its PMT can be seen in Figure 7. The veto array, Figure 8, surrounds the LS volume and is capped by 4 PMTs on the top and the bottom.

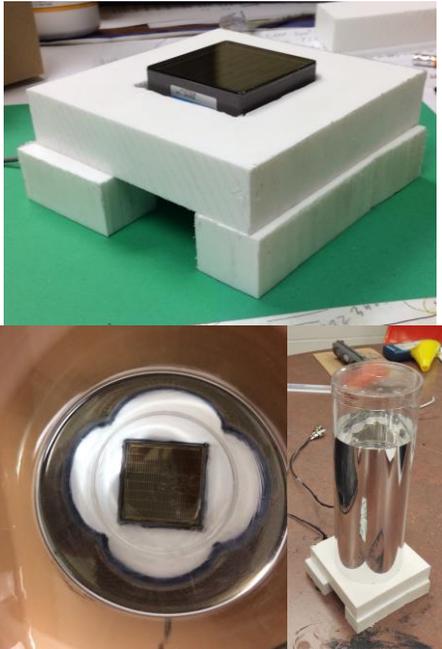

Fig. 7. PMT mounting for liquid scintillator volume (top), mounted LS volume from above (left), and mounted LS from side-on (right).

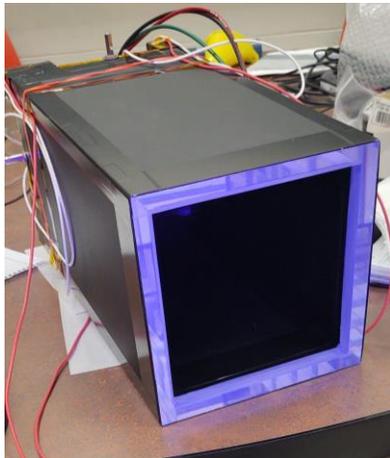

Fig. 8. Veto Array laying horizontally while being prepared for sealing. Note the blue glow due to scintillation from a UV light source incident on the detector.

To test the veto array performance, we ran in two triggering modes. The first set of triggers used a scintillating paddle in double-coincidence with the LS volume to ensure that there was a cosmic ray passing through the LS volume and veto array. Later, we switched to single-coincidence self-triggering on only the LS volume. This data was parsed by a c++ program which searched for failures of the top set or bottom set of PMTs. The vetoing system had an electronics board that would read in the set of four upper PMTs and produce a Boolean signal for vetoing, and similar for the bottom four PMTs.

Later, we replaced the PMTs with a set of six SiPMs. Three of the four corners that had been covered by the PMTs each had a SiPM, and the fourth opening was covered by a reflective layer. The reflective layers were on opposing corners for the top and bottom of the veto array. We chose to only cover three of the four corners because we were using spare SiPMs from the science detector, and there were not enough to cover all eight corners. These SiPM arrays were operated with a DC power supply. The top three and bottom three SiPMs were added as analogue signals for vetoing in software after analysis. Otherwise, the tests performed on the SiPMs were similar to those of the PMT vetoing.

Once the data was collected, the vetoing testing had two analyses to execute. One analysis on the digital signal coming from the FEE board for the veto array, and one analysis on the analog signal coming from the SiPMs. Analyzing the digital signals is very straightforward. A program can quickly look for the timing where the LS PMT gets above a threshold for measurement and look for when the two veto signals come above 50% of their signal height. The time difference between the raw PMT signal and the veto signals can be measured, and then the time window for vetoing can be set. If we see a veto signal within the proper timing window, we can call that a vetoing success, and otherwise a failure for the vetoing set (top or bottom) that failed to fire. When both veto signals fail to trigger with the LS signal, we call this a veto failure.

Analyzing the analog SiPM signals is less straightforward. Instead of having a simple Boolean signal prepared, we must analyze the data by choosing ways of determining veto success or failure. Currently, two methods of determining veto success or failure have been implemented. The two methods currently implemented have been chosen for their simplicity to code and to create outer constraints on the region where the failure rate could be. The method which tends to overestimate the failure rate looks for a veto signal peak within ±40 ns of the LS signal peak. If the peak lies outside that 80 ns window, it is considered to be a veto failure. The optimistic vetoing analysis only looks for a peak in the veto signal where the noise in the measurement is no more than 1/e times the peak value, but the analysis does not have any time constraints other than that the be inside the veto window at all. This amounts to a veto window of 4 μs wide. Further vetoing schema are planned, including using the time at which the veto signal rises above some threshold, and then generates a veto keep out zone approximately 25 ns before the rise and approximately 200 ns after.

### 2.2.2 CubeSat Science Payload Characterization

Radioactive source testing of the GAGG volume has used two types of sources. We have performed tests with γ-ray sources to measure the energy resolution and response of the GAGG-SiPM system with operating voltage and γ energy. We have also used β-decaying elements to verify that this energy characterization allows us to match the spectra of other data sets with regards to the β spectrum. Because of their short penetration range, we did not use any alpha sources with the GAGG crystal for characterization. The signals from the radioactive testing were digitized to csv files, and parsed using a peak-finding algorithm to

To characterize the signals after they have been collected, we used ROOT, a graphing and fitting library maintained by CERN, in combination with c++ code to parse the signals. The c++ code searches through the set of csv files output by the oscilloscope. In each signal trace, the program searches for the peak voltage attained, and puts this value into a histogram of the peak voltages using ROOT. Once the data set has been binned into the histogram, we use the fitting functionality of ROOT to fit Gaussian profiles to the expected locations of the gamma ray corresponding to its energy. These peaks are then used to fit

Additionally, Mark Christl of the NASA Marshall Space Flight Center is currently preparing for similar characterization of the veto array. Once both the veto array and the GAGG detector are characterized, we will perform characterization of the veto array performance with cosmic rays while both shielded and unshielded by the tungsten shield. After cosmic ray testing, the assembly will undergo a final sealing for testing in the Fermilab beamline. The beamline test will characterize high energy protons, electrons, and muons from 800 MeV to 5 GeV.

Detector characterization will consist of placing the payload at a variety of angles inside the beamline such that the beam intersects the GAGG volume and also takes advantage of the symmetries of the detector. In each orientation, we will take large data runs to provide sufficient statistics that we will not want for another beamline test with any of the charged particles we can fire.

## 3. Theory and Calculation

### 3.1 Expected Cosmic Ray Flux Calculations

In order to verify the performance of the CubeSat is as expected, I calculated the expected cosmic ray flux which will pierce the shielding if coming head-on at the thinnest point of the shield. The calculation of this punch through rate was performed in Geant4 where a punch through was defined as any primary or daughter particle surviving to deposit energy on the far side of the shield.

This punch through rate was then used to modulate a spectrum of galactic cosmic rays [3]. The original and modulated spectra for hydrogen during solar minimum can be seen in Figure 10. The full spectrum for all charged nuclei can be approximated by increasing the rate by 10% for heavier nuclei. When the shielded spectrum is calculated, we find that there is a worst-case total flux of 3680 protons / m^2 sr s during solar minimum. This corresponds to approximately 480 protons/s incident on the shield, 125 protons/s incident on the veto array, and 18.1 protons/s incident on the GAGG crystal. Given the low rate of these cosmic ray particles, our expectation is that our fast electronics, ~500 ns, will be able to properly veto the kilohertz-rate signals that the cosmic rays can produce.

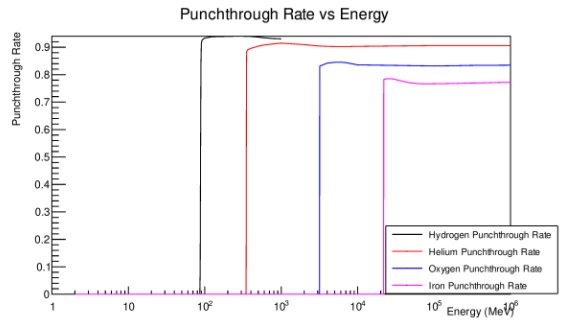

Fig. 9. Cosmic ray spectrum for four elements. Raw rates are shown in black, with shielded rates in red. Spectrum is from solar minimum, when the total cosmic ray flux is at its maximum due to the weaker magnetic shielding from the sun.

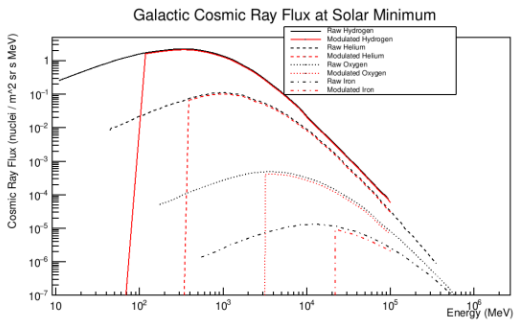

Fig. 10. Cosmic ray spectrum for four elements. Raw rates are shown in black, with shielded rates in red. Spectrum is from solar minimum, when the total cosmic ray flux is at its maximum due to the weaker magnetic shielding from the sun.

### 3.3 Fermilab Beamline Simulation

In preparation for testing in the Fermilab beamline, an undergraduate member of the project has been performing simulations in Geant4. These simulations have iterated over dozens of rotational and beamline energy configurations to prepare the science volume for the eventual beamline test. This simulation uses a volume matching the planned detection, veto, and shielding volumes as described in section 2.1. Some histograms produced by the simulations can be seen below in Figures 11-15 with various particles, energies, and directions. The most noteworthy feature of the simulation is the difference between the electron signal and the proton signal. As we would expect, the electron produces showering, whereas the proton usually acts as a MIP. This leads to a much broader electron spectrum and an order of magnitude higher photon count in comparison to the protons. It is a common feature of the simulation that the rotation generally increases the average signal. The initial position of the pencil beam is directed towards the thinnest part of the GAGG volume, and so each rotation creates a longer path for the charged particles through the scintillator. We also see broader distributions in the histograms because at off-axis angles there is less symmetry in the pencil beam, and there can be much larger variation in the path length, especially for lower-energy particles showering off the primary particle.

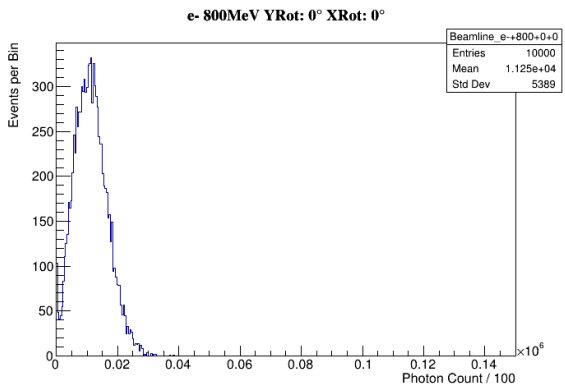

Fig. 11. Electron pencil beam targeted at GAGG volume in simulated detector. Beam is operated at 800 MeV, with no rotation.

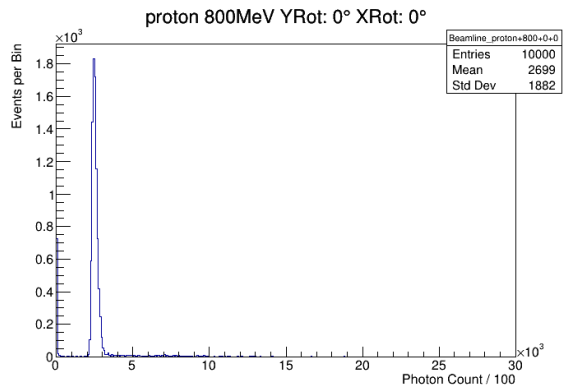

Fig. 12. Proton pencil beam targeted at GAGG volume in simulated detector. Beam is operated at 800 MeV, with no rotation.

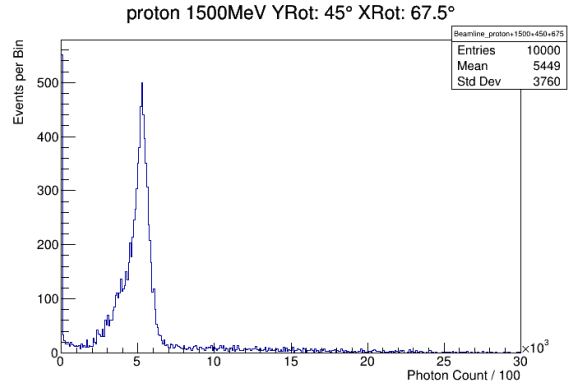

Fig. 13. Proton pencil beam targeted at GAGG volume in simulated detector. Beam is operated at 1500 MeV, with the detector rotated 45 degrees about the global y-axis, and then 67.5 degrees about the global x-axis.0

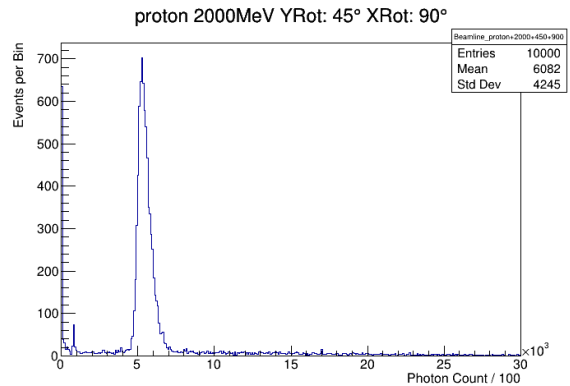

Fig. 14. Proton pencil beam targeted at GAGG volume in simulated detector. Beam is operated at 2000 MeV, with the detector rotated 45 degrees about the global y-axis, and then 90 degrees about the global x-axis.

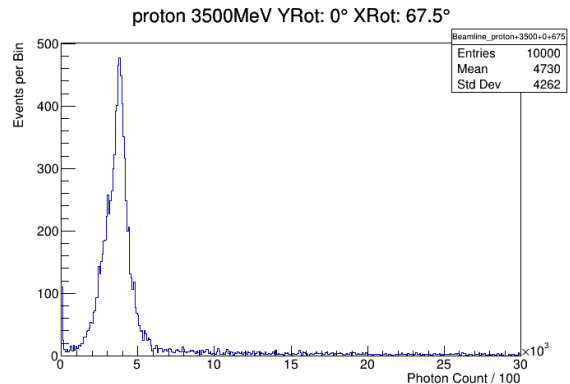

Fig. 15. Proton pencil beam targeted at GAGG volume in simulated detector. Beam is operated at 3500

MeV, with the detector rotated 67.5 degrees about the global x-axis.

## 4. Results and Future Work

### 4.1 Science Volume Calibration

An example of the gamma ray data fitting can be seen in Figure 16. The parameters of this fit are used to find the energy resolution of the detector and to generated data for fitting a calibration curve. After this data had been generated, we have found the curve for each PMT as fitted to the 2-d fit described in (3), where $V_{in}$ is the input voltage of the SiPM, E is the energy in keV, V is the output voltage in mV, and a, b, and c are the fitting parameters.

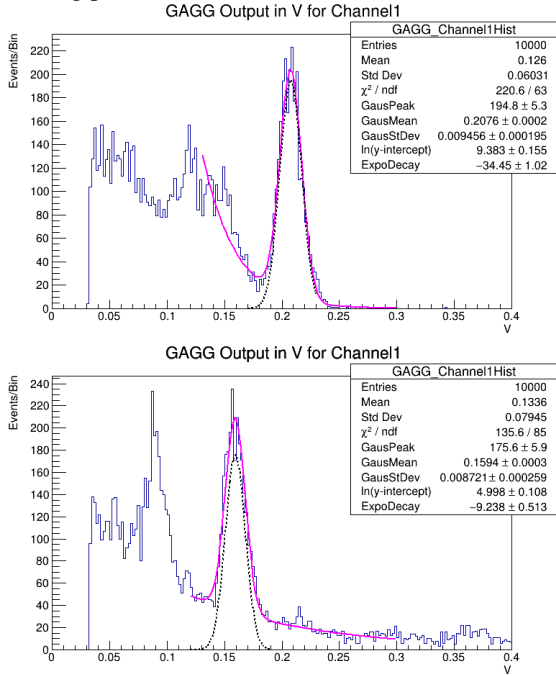

Fig. 16. Fits to gamma ray data for the Cs-137 decay (top) and Na-22 annihilation peak (bottom).

$$V = aV_{in}E + bE + c \qquad (3)$$

The parameters of Eq (3) are listed in Table 2. Of note is that the pedestal of the data acquisition system is consistently low across the operating voltages, and this low threshold is consistent with 0 for each SiPM.

Table 2. Fit parameters of Eq (3). Uncertainties listed are 1 σ.

| SiPM | a | b | c |
|---|---|---|---|
| 1 | 0.0404 ±0.0026 | -2.156 ±0.144 | -2.42 ±4.70 |
| 2 | 0.0132 ±0.0010 | -0.708 ±0.059 | 2.00 ±1.95 |
| 3 | 0.0142 ±0.0009 | -0.760 ±0.053 | 1.62 ±1.74 |
| 4 | 0.0122 ±0.0009 | -0.657 ±0.051 | 1.57 ±1.70 |

### 4.2 Vetoing Rate

The veto failure rate, as described in Section 2.2.1, can be seen graphed in Figure 17 We find that the veto failure rate goes down with energy for the digital PMT vetoing schema and for the optimistic vetoing schema with the SiPMs, as we would expect. Unexpectedly, the failure rate for the 80 ns peak search grew with higher LS thresholds. We expect that this arises from the topology of the higher threshold signals. As the threshold was raised, a larger and larger fraction of the signals results from multiparticle showers, rather than from a single particle passing through. As there are more showers two phenomena arise. First, the rise time of the signals becomes larger, and this could cause signals that were close together in time to drift enough further apart that they are no longer inside a 40 ns window. Second, since the program only looks for the largest peak, not any peak, within 40 ns, the probability that the highest veto peak is within 40 ns decreases. Instead, the highest veto peak becomes likely to correspond to a region of the LS signal other than the highest LS signal value. For example, one of the multiple showering particles might interact strongly in the veto array at a time when there is a relatively small signal in the LS from that or another particle of the shower.

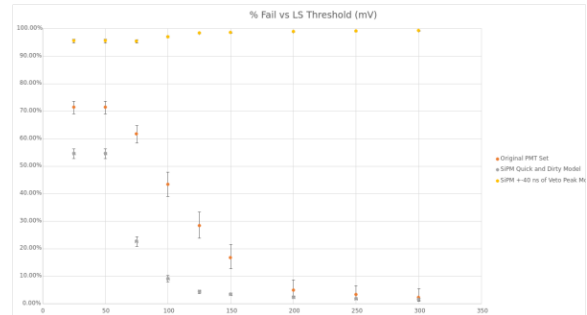

Fig. 17. Veto failure rate vs LS threshold. Three cases are graphed: Failure rate of the original apparatus using the LS, and digital signals coming from the top set of 4 PMTs and the bottom set of PMTs (orange). Failure rate of an aggressive vetoing model on the SiPMs, treating any vetoing signal within the collected data, ~2 μs of the rising edge of the LS, as a valid veto (grey). And a conservative vetoing schema on the SiPMs, where a veto is only considered valid if its peak value is within 40 ns of the LS peak value (yellow). All error bars are 5σ.

### 4.3 Planned Work

Work is currently being done to test a flight-ready veto array, and once this is done, the veto array will be sealed around the GAGG volume. With the completed veto array, we will repeat the testing performed on the test veto and LS volume. After running that test, we will put the passive shield around the veto array and run the test again to see the difference between the shielded and unshielded performance. Once we have finished all of the low-energy characterization for the detector, we will seal the veto array permanently into the passive shielding in preparation for the beamline test. The beamline testing will be performed, the science payload will be stored until the CubeSat arrives from NanoAvionics for mounting. Concurrently, ongoing work with NanoAvionics is being done to finalize and fabricate the design of our CubeSat frame, controller, and other non-science devices.

## 5. Conclusion

In conclusion, work on a CubeSat to test a double-coincidence vetoing schema in real cosmic ray backgrounds is progressing well. The science detector is well-characterized with γ rays, and we are able to find the peak of the strontium-90 β decay peak with good agreement to published results using our GAGG volume. Our GAGG crystal has acceptable energy resolution of 20-30% throughout the 100 keV-1 MeV range, and a test veto array has been able to successfully veto higher energy particles with excellent sensitivity, and lower energy signals will have difficulty penetrating the passive shielding of the detector. Future beamline studies at the Fermilab test beam will help characterize the crystal and veto further in higher energy regimes with protons, electrons, and muons.

Simulations of the expected background rates incident on a shield of identical composition show that we can expect approximately 500 particles per second which will pass through the shield into the veto array or GAGG volumes. Fast electronics will be able to handle such a rate, and this will provide an excellent sample of data for analyzing the veto array's failure rate in-situ. These results will help to raise the technological readiness of our project to produce a similar detector schema for a near-solar mission to demonstrate neutrino detection in space.

## Acknowledgements

I would like to thank Dr. Nick Solomey for creating and leading the vSOL project and the NASA Advanced Innovative Concepts office for funding our project through three phases of research culminating in the building of this CubeSat. Special thanks to Jarred Novak for his tireless engineering for the project and for his work on designing and simulating the doped-epoxy shielding. I would like to thank Dr. Mark Christl for designing, testing, and fabricating the veto arrays for the project. Thanks to Brian Doty, Trent English, and Octavio Pacheco-Vazquez for their help with collecting the GAGG calibration data, analyzing said data, and simulating both the GAGG volume and beamline tests.